\documentclass[a4paper, draftcls, onecolumn]{IEEEtran}
\ifCLASSOPTIONcompsoc
\else
\fi
\ifCLASSINFOpdf
\else
\fi
\usepackage{latexsym}
\usepackage[fleqn]{amsmath}
\usepackage{amssymb}
\usepackage{graphicx}

\newtheorem{theorem}{Theorem}
\newtheorem{corollary}{Corollary}

\newtheorem{remark}{Remark}

\newtheorem{code}{Coding Scheme}
\newcommand{\bK}{\mbox{\boldmath $K$}}
\newcommand{\bsK}{\mbox{\boldmath ${\scriptstyle K}$}}

\begin{document}
\title{Multiple Object Identification Coding}

\author{Hirosuke Yamamoto,~\IEEEmembership{Fellow,~IEEE,}
        Masashi Ueda% <-this % stops a space
\IEEEcompsocitemizethanks{\IEEEcompsocthanksitem H.Yamamoto is with the Department of Complex Science and Engineering, 
The University of Tokyo, Kashiwa-shi, Chiba, 277-8561 Japan.
\protect\\
E-mail: hirosuke@ieee.org
\IEEEcompsocthanksitem M.~Ueda was with the Department of Mathematical Informatics, The University of Tokyo, Bunkyo-ku, Tokyo,113-8656 Japan.
}% <-this % stops a space
\thanks{This work was presented in part at the IEEE ISIT2014. 
This work was supported in part by JSPS KAKENHI Grant Numbers 26630169 and 25289111.
}}

\IEEEtitleabstractindextext{%
\begin{abstract}
In the case of ordinary identification coding,
a code is devised to identify a single object among $N$ objects. 
But, in this paper, we consider an identification coding problem to 
identify $K$ objects at once among $N$ objects  in the both cases 
that $K$ objects are ranked or not ranked.
By combining Kurosawa-Yoshida scheme with Moulin-Koetter scheme, 
an efficient identification coding scheme is proposed, which can attain high coding rate and error exponents compared with the case that an ordinary identification code is used $K$ times. Furthermore, the achievable triplet of rate and error exponents of type I and type II 
decoding error probabilities are derived for the proposed coding scheme.
\end{abstract}

}

\maketitle

% is not selected <OR> if conference mode is selected - because compsoc
% conference papers position the abstract like regular (non-compsoc)
\IEEEdisplaynontitleabstractindextext

% \fi
\IEEEpeerreviewmaketitle

\section{Introduction}
Consider a case such that we must inform many receivers about a winner, who is selected among them, via a stationary discrete memoryless channel.  If each receiver is interested only in whether he/she is the winner or not, but is not interested in who wins when he/she is not the winner, an identification code (ID code) can be used to transmit the information efficiently.
It is known that the decoding error probability of each receiver can become arbitrarily small if $R<C$,  where $C$ is the channel capacity and $R$ is the coding rate of the ID code defined by $R=(\log\log N)/n$ for the number of receivers $N$ and the code length $n$ \cite{ad1989}\cite{hv1992}.

Verd\'{u} and Wei \cite{vw1993} showed that 
an ID code for a noisy channel can be constructed by concatenating an ID code for the noiseless channel and a transmission code (an ordinary error correcting code) for the noisy channel. They also gave an ID code for the noiseless channel by using a constant weight matrix based on Reed-Solomon codes.
Furthermore, Kurosawa and Yoshida \cite{ky1999} showed that a more efficient ID code for the noiseless channel can be constructed by using $\varepsilon$-almost strongly universal
classes of hash functions, and Moulin and Koetter \cite{mk2006} proposed another construction scheme of ID codes based on Reed-Solomon codes, which is efficient if common randomness can be used among the sender and receivers.

In this paper, we consider the case that there are $K$ winners among $N$ receivers. In this case, we can send the information of winners by using an ordinary ID code $K$ times. But, the coding rate is decreased to $R/K$. If we construct an ordinary ID code for $\tilde N=\binom{N}{K}$ and assign $\binom{N-1}{K-1}$ indices to each receiver, we can send the information with the same coding rate $R$ as the case of $K=1$. However, the type II decoding error probability becomes very large because each receiver must decode the received word for all $\binom{N-1}{K-1}$ indices. This means that  
the type II decoding error probability becomes $\binom{N-1}{K-1}$ times as large as the case of $K=1$. 

We note that Ahlswede \cite{ahlswede2006}\cite{ahlswede2008} studied {\em $K$-Identification}. Let ${\cal N}$ and ${\cal K}_i$ be the set and a subset of all receivers, respectively, where $|{\cal N}|=N$ and $|{\cal K}_i|=K$, and $|\cdot|$ represents the cardinality of a set. 
Then, it is assumed in the $K$-identification problem that each receiver $i$ knows the set ${\cal K}_i$, a codeword is encoded from only one $\hat{i}\in {\cal N}$, and each receiver $i$ wants to know whether $\hat{i}\in{\cal K}_i$ or $\hat{i}\not\in{\cal K}_i$. 
In \cite{abk2006}, the $K$-Identification is further generalized to {\em Generalized Identification}, in which each receiver $i$ not only finds out whether $\hat{i}\in{\cal K}_i$ or $\hat{i}\not\in{\cal K}_i$, but also identifies $\hat{i}$ if $\hat{i}\in{\cal K}_i$.
But, it is still assumed in the Generalized Identification that each receiver $i$ knows ${\cal K}_i$ and a codeword is encoded from only one $\hat{i}\in {\cal N}$. In contrast, we assume in our coding problem that any receiver doesn't know ${\cal K} (\subset {\cal N})$, which is the set of winners selected at the sender side, a codeword is encoded from ${\cal K}$, and each receiver $i$ wants to know whether $i\in{\cal K}$ or $i\not\in{\cal K}$.
So, since our coding problem is quite different from $K$-Identification and Generalized Identification,  we cannot use  their coding schemes for our coding problem.

We call our identification coding problem Multiple Object Identification (MOID) to distinguish from $K$-Identification and Generalized Identification.

In this paper, we show that an efficient explicit MOID code can be constructed by combining Kurosawa-Yoshida coding scheme  \cite{ky1999} and Moulin-Koetter coding scheme \cite{mk2006}. We derive the achievable region of coding rate and exponents of type I and type II decoding error probabilities. 
In Sections 2 and 3, we treat the cases that $K$ winners are not ranked and are ranked,
respectively.

For simplicity we first assume that $K$ is fixed. But the case of variable $K$ is considered in Section \ref{sec-2-F}.  Furthermore,  in Sections \ref{sec-2-D} and \ref{sec-2-E},  we treat the cases  that 
the noiseless feedback channel and common randomness can be used between the sender and receivers.
An ordinary error correcting code is called a transmission code to distinguish from an ID code in this paper, and the combined MOID coding with transmission coding is treated in Section \ref{sec-2-C}.

\section{MOID code without ranking}\label{sec-2}
\subsection{Definition of MOID codes}
Let ${\cal N}\equiv\{1, 2, \cdots, N\}$ be the set of objects 
and let ${\mathcal  K}$ be a subset of  ${\cal N}$, which is selected 
at the sender side. For simplicity,  {\em objects} are called {\em receivers} in the following.

The sender sends binary information $u_i\in{\cal U}\equiv\{{\rm T}, {\rm F}\}$ 
to each receiver $i$ such that $u_i={\rm T}$ if $i\in {\cal K}$ and $u_i={\rm F}$ if $i\not\in {\cal K}$.
In other words, ${\cal K}$ can be represented as follows.
\begin{equation}
{\cal K}\equiv \{ i : u_i= {\rm T}, i\in {\cal N}\}, \label{eq-1}
\end{equation}
For simplicity, we assume  that $K\equiv|{\cal K}|\geq 1$ is fixed. Let ${\cal Z}\equiv \{{\cal K}\}$ be the set of all possible ${\cal K}$. Then we note that 
$|{\cal Z}|$ is given by $\binom{N}{K}$, and the ordinary ID coding corresponds to the case of $K=1$. 

The channel is a discrete memoryless channel (DMC) $W$ with input alphabet ${\cal X}$
and output alphabet ${\cal Y}$. For simplicity, we assume that the channel input is binary, i.e.~$|{\cal X}|=2$. But, the results can easily be extended to the case of $|{\cal X}|\geq 2$. 
We also assume that the encoder $\varphi$ of MOID code can use a random number $v$ which takes a value of ${\cal V}=\{1, 2, \cdots$, $|{\cal V}|\}$. Then, the encoder $\varphi$  to identify $K$ receivers can be defined as follows.
\begin{equation}
\varphi : {\cal Z} \times {\cal V} \rightarrow {\cal X}^n, \label{eq-2}
\end{equation}
where $n$ is the code length, and a codeword $x^n$ is generated by $x^n=\varphi({\cal K}, v)$
from MOID information ${\cal K}\in{\cal Z}$ and random number $v\in {\cal V}$.
This means that the encoder $\varphi$ is a stochastic encoder for a given ${\cal K}$.
The decoder $\psi_i$ of receiver $i$, which outputs T or F, is defined as follows.
\begin{equation}
\psi_i : {\cal Y}^n \rightarrow  {\cal U}. \label{eq-3}
\end{equation}
An MOID code $(\varphi, \psi_1, \psi_2, \cdots, \psi_N)$ is called a $K$-MOID code
if $K=|{\cal K}|$. 

The coding rate $R_K^{(n)}$ of a $K$-MOID code is defined by\footnote{The base of logarithm is always 2 in this paper.}
\begin{equation}
R_K^{(n)}\equiv\frac{1}{n} \log\log N. \label{eq-4}
\end{equation}

Next we consider the decoding error probabilities of a $K$-MOID code.
Type I decoding error probability and its exponent are defined as follows.
\begin{align}
 \lambda_1^{(n)}(i|{\cal K})&\equiv{\rm Pr}\{\psi_i(\varphi({\cal K}, V))={\rm F}\} \quad\mbox{for $i\in {\cal K}$},\label{eq-5}\\
 \lambda_1^{(n)}& \equiv\max_{{\cal K}\in {\cal Z}}\max_{i\in {\cal K}} \lambda_1^{(n)}(i|{\cal K}), \label{eq-5-1}\\
 E_1^{(n)}& \equiv-\frac{1}{n}\log \lambda_1^{(n)}, \label{eq-5-2}
\end{align}
where $\lambda_1^{(n)}(i|{\cal K})$ represents the decoding error probability of receiver $i\in {\cal K}$, $\lambda_1^{(n)}$ is the worst of $\lambda_1^{(n)}(i|{\cal K})$,
and $E_1^{(n)}$ is the exponent of $\lambda_1^{(n)}$.

Similarly,  type II decoding error probability is defined by
\begin{align}
 \lambda_2^{(n)}(i|{\cal K})&\equiv{\rm Pr}\{\psi_i(\varphi({\cal K}, V))={\rm T}\}\quad\mbox{for $i\not\in {\cal K}$}, \label{eq-6}\\
 \lambda_2^{(n)}& \equiv\max_{{\cal K}\in {\cal Z}}\max_{i\not\in {\cal K}} \lambda_2^{(n)}(i|{\cal K}), \label{eq-6-1}\\
  E_2^{(n)}& \equiv-\frac{1}{n}\log \lambda_2^{(n)}, \label{eq-6-2}
\end{align}
where $\lambda_2^{(n)}(i|{\cal K})$ is the decoding error probability of receiver $i\not\in {\cal K}$,
$\lambda_2^{(n)}$  is the worst of $\lambda_2^{(n)}(i|{\cal K})$, and $E_2^{(n)}$ is the exponent of $\lambda_2^{(n)}$. 

A triplet $(R, E_1, E_2)$ is said to be achievable by a coding scheme if the following inequalities can be satisfied by the coding scheme. 
\begin{align}
\liminf_{n\rightarrow \infty} R_M^{(n)}&\geq R\label{eq-7}\\
\liminf_{n\rightarrow \infty} E_1^{(n)} &\geq E_1 
\label{eq-8}\\
\liminf_{n\rightarrow \infty} E_2^{(n)} &\geq E_2 
\label{eq-9}
\end{align}

\begin{remark}\label{remark-1}
When $K=1$, the $K$-MOID code coincides with the ordinary ID code, and coding rate $R_K^{(n)}$ and error exponents $E_1^{(n)}$ and $E_2^{(n)}$  also coincide with the ones of the ordinary ID code.
\end{remark}

For $K=1$, the following triplet  is achievable by Verd\'{u}-Wei coding scheme \cite{vw1993} and Kurosawa-Yoshida coding scheme \cite{ky1999}.
\begin{align}
(R, E_1, E_2)=&\left(\left(1-\frac{3}{\ell}\right) r, E(r), \min\left\{\frac{r}{\ell}, E(r)\right\}
\right),  \nonumber\\
&\hspace{0.5cm}0 < r < C, \quad\ell = 3, 4, 5, \cdots,  \label{eq-10}
\end{align}
where $E(r)$ is the reliability function (or the error exponent)  of DMC $W$ in transmission coding,  $C$ is the capacity of $W$ given by $C=\max_{P_X} I(X; Y)$, and $r$ and $\ell$ are parameters that we can select freely.
Furthermore, the following triplet is also achievable by Verd\'{u}-Wei coding scheme \cite{vw1993} and Moulin-Koetter coding scheme \cite{mk2006}.
\begin{align}
(R, E_1, E_2)=&(\rho r, E(r), \min\{(1/2 -\rho)r, E(r)\}), \nonumber\\
&\hspace{0.5cm}0 < r < C, \quad 0\leq \rho \leq1/2, \label{eq-11}
\end{align}
where $r$ and $\rho$ are parameters.

We note from \eqref{eq-10} that we can attain $\displaystyle{\lim_{n\rightarrow\infty} \lambda_1^{(n)}=0}$
and $\displaystyle{\lim_{n\rightarrow\infty} \lambda_2^{(n)}=0}$
for any $0<R<C$ by setting $r$ sufficiently close to $C$ and $l$ sufficiently large.

\vspace{0.2cm}
\subsection{Construction of MOID codes}
We construct an MOID code for a noisy channel by cocatinating  an MOID code for the noiseless channel and a transmission code for the noisy channel in the same way as 
\cite{vw1993}.

We first review the known coding schemes for the noiseless channel in the case of $K=1$, i.e.
the ordinary  ID coding.  In Verd\'{u}-Wei scheme \cite{vw1993} and Kurosawa-Yoshida scheme \cite{ky1999}, a codeword of ID information $i$ is given by a random number $v$, which is distributed uniformly over a subset ${\cal V}_i\subset {\cal V}$. The subset ${\cal V}_i$ depends on $i$ and is determined based on Reed-Solomon code in \cite{vw1993} or based on $\varepsilon$-almost strongly universal classes of hash functions in \cite{ky1999}.  
These coding schemes can be extended to the MOID coding by replacing a single $v$ with a $K$ dimensional vector $(v_1, v_2, \cdots, v_K), v_j\in{\cal V}_{i_j} \subset {\cal V}$ for ${\cal K}=\{i_1, i_2, \cdots, i_K\}$. 
But, since the code length becomes $K$ times long,  
the coding rate decreases to $1/K$. 
On the other hand, the codeword of ID information $i$ consists of $(v, c_v(i))$ in Moulin-Koetter scheme \cite{mk2006}, where  $c_v(i)$ is constructed based on Reed-Solomon code.
Their scheme can be extended to the MOID coding by replacing the codeword with $(v, c_v(i_1), c_v(i_2), \cdots, c_v(i_K))$. But, since $v$ and $c_v(i)$ 
must satisfy  $\|v\|= \|c_v(i)\|$ in their scheme, where $\|a\|$ represents the bit length of $a$, 
the code length becomes $(K+1)/2$ times longer and the coding rate decreases to $2/(K+1)$. Hence, the above extensions of known schemes are inefficient for the MOID coding.

Instead of $(v, c_v(i))$, we use a codeword $(v, h_v(i))$, where $c_v(i)$ is replaced with a hash function $h_v(i)$ satisfying that $\|v\|\gg \|h_v(i)\|$. 
In this case, even if we extend the codeword to $(v, h_v(i))$ to $(v, h_v(i_1), h_v(i_2), \cdots, h_v(i_K))$ for the MODI coding, the coding rate does not decrease significantly.

Now we describe our coding scheme for the MOID coding.
We use the same $\varepsilon$-almost strongly universal classes of hash functions ${\cal H}=\{h_l \}$ as Kurosawa-Yoshida scheme \cite{ky1999}, which satisfies 
the following relations for  $h_l : \mathcal{A}\rightarrow \mathcal{B}$.
\begin{align}
 |\{ h_l\in\mathcal{H} : & h_l(\alpha)=\beta\}|=\dfrac{|\mathcal{H}|}{|\mathcal{B}|}\nonumber\\
 &\hspace*{1.5cm} \mbox{for $\forall \alpha\in\mathcal{A},\forall\beta\in\mathcal{B}$}
 \label{eq-10-b}\\
|\{ h_l\in\mathcal{H} : &h_l(\alpha_1)=\beta_1,h_l(\alpha_2)=\beta_2\}|\leq \varepsilon \dfrac{|\mathcal{H}|}{|\mathcal{B}|} \nonumber\\
&\hspace*{-1cm}\mbox{for $\forall \alpha_1,\alpha_2\in\mathcal{A}, \alpha_1\neq \alpha_2,\forall \beta_1,\beta_2\in\mathcal{B}$ }
\label{eq-11-b}
\end{align}

In order to construct a $K$-MOID code, we set ${\cal A}$ and ${\cal H}$ as ${\cal A}={\cal N}$
($|{\cal A}|=N$) and $|{\cal H}|=|{\cal V}|$, respectively.
Let $f$ and $g$ be the encoder and decoder, respectively, of a transmission code for noisy channel $W$ such that $f: {\cal V}\times \beta^K \rightarrow {\cal X}^n$ and $g: {\cal Y}^n\rightarrow {\cal V}\times \beta^K$.
Then,  we construct $K$-MOID code $(\varphi, \psi_1, \psi_2, \cdots, \psi_N)$ as follows.
\vspace{0.2cm}
\begin{code}\label{code-1}
\begin{align}
&\mbox{Encoder $\varphi$ :} \nonumber\\
&\quad\mbox{For ${\cal K}=\{i_1, i_2, \cdots, i_K\}\subset {\cal N}$},\nonumber\\
&\quad\quad\varphi({\cal K}, v) \equiv f(v, h_v(i_1), h_v(i_2), \cdots, h_v(i_K)). \label{eq-12}\\
&\mbox{Decoder $\psi_i$:} \nonumber\\
&\quad\quad\psi_i(y^n)
\equiv\left\{
\begin{array}{ll}
{\rm T}, & \mbox{if $h_{\hat{v}}(i)=\beta_j$ holds }\\
& \quad\quad  \mbox{for some $j$, $1\leq j \leq K$} \\
{\rm F}, & \mbox{otherwise}
\end{array}\right.\nonumber \\
&\hspace{1.5cm}\mbox{for $(\hat{v}, \beta_1, \beta_2, \cdots, \beta_K)= g(y^n)$}, \label{eq-13}
\end{align}
where $v$ is a random number distributed uniformly over ${\cal V}$.
\end{code}

This $K$-MOID code satisfies the following theorem.

\vspace{0.2cm}
\begin{theorem}\label{thm-1}
The following triplet is achievable by Coding Scheme 1.
\begin{align}
(R, &E_1, E_2)=\nonumber\\
&\left(\left(1-\frac{K+3}{K+\ell}\right) r, E(r), \min\left\{\frac{r}{K+\ell}, E(r)\right\}
\right),  \nonumber\\
&\hspace{1cm}0 <r < C, \quad\ell =3, 4, 5, \cdots. \label{eq-14}
\end{align}
\end{theorem}

\vspace{0.2cm}
{\it Proof} \quad
First we construct a $K$-MOID code with code length $n_0$ for the binary noiseless channel. 

We use the above $\varepsilon$-strongly universal classes of hash functions. 
Setting $n_0=q^k$ and $d=q^k-q^t+1$ in \cite[Corollary 3.1]{ky1999}, 
we have for $q=2^m$ that 
\begin{align}
|{\cal A}|&= N= q^{kq^t},      \label{eq-15}\\
{\cal B}&= {\rm GF}(q)\quad (|{\cal B}| = q ), \label{eq-16}\\
|{\cal V}|&=|{\cal H}|=q^{k+2},  \label{eq-17}\\
\varepsilon& = \frac{k}{q}+\frac{q^t-1}{q^k}
\leq \frac{1}{q}\left(k+\frac{q^t}{q^{k-1}} \right), \label{eq-18}
\end{align}
where $t\leq k-1$ because it must hold that $\varepsilon\rightarrow 0$
as $m\rightarrow \infty$ (i.e., $q\rightarrow \infty$).

Then, from \eqref{eq-16}, \eqref{eq-17}, and $q=2^m$, 
the code length $n_0=\|(v, h_v(i_1), h_v(i_2), \cdots, h_v(i_K))\|$
is given by
\begin{align}
n_0=\log|{\cal V}| +K \log |{\cal B}|  = (k+2+K)m. \label{eq-19}
\end{align}
Hence, from \eqref{eq-15} and
\eqref{eq-19}, the coding rate of this code satisfies 
\begin{align}
R_K^{(n_0)} &= \frac{1}{n_0}\log\log N\nonumber\\
&=\frac{1}{n_0}\log\left\{kq^t\log q\right\}\nonumber\\
&=\frac{1}{n_0}\left\{tm+\log k+\log m\right\}\nonumber\\
&=\frac{t}{k+2+K}+\frac{1}{n_0}(\log k+\log m) \nonumber\\
&= \frac{t}{k+2+K} + O\left(\frac{\log n_0}{n_0}\right).\label{eq-20}
\end{align}
Since the optimal $t$ that maximizes \eqref{eq-20} for $1\leq t \leq k-1$ is $t=k-1$,
we can attain the following coding rate.
\begin{align}
R_K^{(n_0)} &=  \frac{k-1}{k+2+K} + O\left(\frac{\log n_0}{n_0}\right)\nonumber\\
&=1- \frac{K+3}{k+2+K} + O\left(\frac{\log n_0}{n_0}\right) \label{eq-21}
\end{align}

Next we evaluate the decoding error probabilities. 
In the case of the noiseless channel, every $\psi_i$ always outputs T if $i\in {\cal K}$. Hence
for any ${\cal K}\in{\cal Z}$ and any $i\in {\cal K}$, $\lambda_1^{(n_0)}(i|{\cal K})=0$. This means that
$\lambda_1^{(n_0)}=0$ and $E_1^{(n_0)}=\infty$.

For ${\cal K}=\{i_1, i_2, \cdots, i_K\}$ and $i\not\in {\cal K}$,  $\lambda_2^{(n_0)}(i|{\cal K})$ is bounded as follows.
\begin{align}
\lambda_2^{(n_0)}(i|{\cal K}) &=\text{Pr}\left\{\bigcup_{j=1}^K \left(h_V(i)=h_V(i_j)\right)\right\}\nonumber\\
&\leq \sum_{j=1}^K 
\text{Pr}\left\{h_V(i)=h_V(i_j)\right\}\nonumber\\
&=K\frac{\sum_{\beta\in {\cal B}}|\{h_v: h_v(i)=h_v(i_j)=\beta\}|}{|{\cal V}|}\nonumber\\
&\leq {\varepsilon}K,  \label{eq-22}
\end{align}
where the first and second inequalities hold from the union bound and \eqref{eq-11-b},
respectively.
Since this bound does not depend on ${\cal K}$ and $i\not\in {\cal K}$, $\lambda_2^{(n)}$ has the same bound.
\begin{equation}
\lambda_2^{(n_0)}\leq  {\varepsilon}K \label{eq-22-2}
\end{equation}
Next we evaluate $E_2^{(n)}$, the exponent of $\lambda_2^{(n)}$.
From \eqref{eq-6-2}, \eqref{eq-18}, \eqref{eq-19}, and \eqref{eq-22-2},
$E_2^{(n_0)}$ has the following bound for $t\leq k-1$.
\begin{align}
\hspace*{-0.4cm}
E_2^{(n_0)}&\geq -\frac{1}{n_0}\{\log K+\log \varepsilon\}\nonumber\\
&\geq -\frac{1}{n_0}\left\{\log K-\log q+\log\left(k+\frac{q^t}{q^{k-1}}\right)\right\}\nonumber\\
&= \frac{1}{ k+2+K}-\frac{1}{n_0}\left\{\log K+\log\left(k+\frac{q^t}{q^{k-1}}\right)\right\}\nonumber\\
&= \frac{1}{ k+2+K}-O\left(\frac{\log k}{n_0}\right)\label{eq-23}
\end{align}

Setting $\ell=k+2$, $\ell=3, 4, \cdots$, and $m\rightarrow \infty$, i.e.~$n_0\rightarrow\infty$,  in \eqref{eq-21} and \eqref{eq-23}, 
we note that the following triplet is achievable for the binary noiseless channel.
\begin{equation}
(R, E_1, E_2)=
\left(1-\frac{K+3}{K+\ell},\; \alpha, \; \frac{1}{K+\ell}\right), \label{eq-24}
\end{equation}
where $\alpha>0$ is an arbitrarily large constant.

Next we treat the case of binary DMC $W$.
If we transmit $(v, h_v(i_1), h_v(i_2), \cdots, h_v(i_K))$ via $W$ by using the best transmission code $(f, g)$ of $W$ with coding rate $r$, $0<r<C$, then the code length $n$ is given by $n=n_0/r$ and the decoding error probability of the transmission code is upper bounded by $2^{-nE(r)}$, where $E(r)$ and  $C$ are the reliability function and the capacity of $W$, respectively.
Hence, the total error probability $\lambda_j^{(n)}$, $j=1, 2$, is bounded as follows.
\begin{align}
\hspace{-0.2cm}
\lambda_j^{(n)} \leq  2^{-n_0E^{(n_0)}_j}+ 2^{-nE(r)}  \leq  2^{-n \min\{rE^{(n_0)}_j, E(r)\}}  \label{eq-25}
\end{align}

From \eqref{eq-24} and \eqref{eq-25}, the triplet given by \eqref{eq-14} is achievable.

\vspace{-0.2cm}
\begin{flushright}Q.E.D.\end{flushright}

\vspace{0.2cm}
\begin{remark}
In \eqref{eq-14}, we have $R=0$ when $\ell=3$. In this case, $R_K^{(n)}\equiv (\log\log N)/n$
tends to zero as $n\rightarrow 0$. But, $\widehat{R}_K^{(n)}\equiv (\log N)/n$ does not tend to zero because it holds from \eqref{eq-20} that for $t=k-1=\ell-3=0$,
\begin{align}
\widehat{R}_K^{(n)} &= \frac{\log N}{n}\nonumber\\
&=\frac{kq^{t}\log q}{n}\nonumber\\
&=\frac{m}{(3+K)m/r} \nonumber\\
&=\frac{r}{3+K}.
\end{align}
Hence, the case of $\ell=3$ is not meaningless.
\end{remark}

\vspace{0.2cm}
\begin{remark}\label{rmk-4}
If we use Verd\'{u}-Wei's ID code or Kurosawa-Yoshida's ID code $K$ times, the following triplet can be achieved from \eqref{eq-10}.
\begin{align}
(R, &E_1, E_2) \nonumber\\
 = &\left(\frac{1}{K}\left(1-\frac{3}{\ell}\right) r, \frac{E(r)}{K}, \min\left\{\frac{r}{\ell K}, \frac{E(r)}{K}\right\}
\right),  \nonumber\\
&\hspace{2.5cm}0 \leq r \leq C, \quad\ell = 3, 4, 5, \cdots \label{eq-26}
\end{align}
If we use $(v, c_v(i_1), c_v(i_2), \cdots, c_v(i_K))$ in Moulin-Koetter scheme,
we can achieve 
\begin{align}
(R, &E_1, E_2)\nonumber\\
 = &\left(\frac{2\rho r}{K+1}, \frac{2E(r)}{K+1}, \min\left\{\frac{(1 -2\rho)r}{K+1}, \frac{2E(r)}{K}\right\}\right), \nonumber\\
&\hspace{2.5cm}0 < r < C, \quad 0\leq \rho \leq1/2. \label{eq-26-2}
\end{align}
We can easily check that \eqref{eq-14} is much better than \eqref{eq-26} and \eqref{eq-26-2} for $K\geq 2$.
\end{remark}

\vspace{0.2cm}
\begin{remark} \label{rmk-1}
From Theorem \ref{thm-1},
Coding Scheme 1 can achieve for $K=1$ that
\begin{align}
(R, &E_1, E_2)
\nonumber\\
= &\left(\left(1-\frac{4}{1+\ell}\right) r, E(r), \min\left\{\frac{r}{1+\ell}, E(r)\right\}
\right),  \nonumber\\
&\hspace{2.5cm}0 < r < C, \quad\ell = 3, 4, 5,  \cdots \label{eq-10-a}
\end{align}
This triplet is a little worse than \eqref{eq-10}. But Coding Scheme 1 can attain high performance for $K\geq 2$. Furthermore, it has advantages for  $K\geq 1$ if the encoder and decoders can use common randomness or a noiseless feedback channel as shown in Sections  \ref{sec-2-D} and \ref{sec-2-E}.
\end{remark}

\vspace{0.2cm}
\begin{corollary}\label{cor-1}
The $K$-MOID code constructed by Coding Scheme 1 can achieve 
\begin{align}
\lim_{n\rightarrow\infty} R^{(n)} =C, \label{eq-c-1}\\
\lim_{n\rightarrow\infty} \lambda_1^{(n)} =0, \label{eq-c-2}\\
\lim_{n\rightarrow\infty} \lambda_2^{(n)} =0. \label{eq-c-3}
\end{align}
\end{corollary}

{\it Proof} \quad
For an arbitrarily given $\xi>0$, we select $r$ and $\ell$ that satisfy the following inequalities.
\begin{align}
C\left(1-\frac{\xi}{2}\right) &< r <C  \label{eq-c-4} \\
\frac{K+3}{K+\ell} &< \frac{\xi}{2} \label{eq-c-5}
\end{align}
Then, for sufficiently large $n$,  coding rate $R_K^{(n)}\approx \left(1-\frac{K+3}{K+\ell}\right)r$ satisfies
\begin{align}
C(1-\xi) < R_K^{(n)} < C. \label{eq-c-6}
\end{align}
From \eqref{eq-c-4}, we have $E(r)>0$. Obviously $\frac{r}{K+\ell}>0$. Hence \eqref{eq-c-2}
and \eqref{eq-c-3} hold because their exponents are positive.
Since the above holds for any $\xi>0$, \eqref{eq-c-1} is obtained by setting $\xi\rightarrow 0$ as $n\rightarrow\infty$.
\vspace{-0.4cm}\begin{flushright}Q.E.D.\end{flushright}

\vspace{0.2cm}
\begin{remark} \label{rmk-0}
In order to attain \eqref{eq-c-1}, $\ell$ must be sufficiently large and $r$ must be sufficiently  close to $C$. This means that $E_1\rightarrow 0$ and $E_2 \rightarrow 0$ even though
\eqref{eq-c-2} and \eqref{eq-c-3} hold.
\end{remark}

\subsection{$K$-MOID Coding with a Transmission Message} \label{sec-2-C}
 \label{rmk-ID-T}
It is shown in \cite{hv1992} that an ID code can send a transmission message in addition to an ID message at once. 
Actually ID codes given by \cite{vw1993}--\cite{mk2006}
can realize such coding. Similarly, Coding Scheme 1 can send a transmission message in addition to a $K$-MOID message at once by replacing the random number $v$ with 
a transmission message which is distributed uniformly over ${\cal V}$.

In this case, the coding rate $R_T^{(n)}$ of the transmission message is given by 
\begin{align}
R_T^{(n)}&\equiv \frac{1}{n} \log |{\cal V}|\nonumber\\
&=\frac{n_0}{n}\frac{1}{n_0}\log|{\cal V}|\nonumber\\
&=r\frac{\ell}{\ell+K}, \quad \ell=3, 4, \cdots
\end{align}
from \eqref{eq-17} and \eqref{eq-19}. Hence, by setting $r$ sufficiently close to $C$ and $\ell$ sufficiently large, we can achieve 
\begin{align}
\lim_{n\rightarrow\infty} R_T^{(n)}=C \quad \mbox{and} \quad 
\lim_{n\rightarrow\infty} P_{Te}^{(n)}=0
\end{align}
in addition to $\displaystyle{\lim_{n\rightarrow\infty} R_K^{(n)}=C}$ and
$\displaystyle{\lim_{n\rightarrow\infty} \lambda_i^{(n)}=0}$, $i=1,2$
 at once, where $P_{Te}^{(n)}$ is the decoding error probability of the transmission message.

\subsection{$K$-MOID Coding with Common Randomness} \label{sec-2-D}

If the encoder and decoders can use common randomness, e.g.~a good  pseudo random number generator, we don't need to send some or all  bits of random number $v$ in the same way as Moulin-Koetter scheme.

Assume that we can use $n_{0c}$ bit common randomness, 
and define the rate of the common randomness by $R_c=n_{0c}/n_0$.
Then, from \eqref{eq-19}, $n_0=(\ell+K)m$ and $0\leq n_{0c}\leq \ell m$ for 
$k+2=\ell =3, 4, \cdots$. 
Since we don't need send $n_{0c}=R_c n_0$ bits, the code length can be shortened to $n_0- R_cn_{0c}=n_0(1-R_c)$ bits. 
This means that achievable $(R, E_1, E_2)$ can be enlarged to $(R/(1-R_c), E_1/(1-R_c), E_2/(1-R_c))$ by using common randomness with rate $R_c$.

Now consider the case of maximum $R_{c}$, i.e.~$R_c=\ell/(\ell+K)$.
In this case, we can attain from \eqref{eq-14} that
\begin{align}
(&R, E_1, E_2)=\nonumber\\
&\left(\frac{(\ell -3)r}{K}, \frac{(\ell+K) E(r)}{K}, \min\left\{\frac{r}{K}, \frac{(\ell+K) E(r)}{K}\right\}
\right),  \nonumber\\
&\hspace{1cm}0 <r < C, \quad\ell =3, 4, 5, \cdots.
\end{align}
Hence, $R$ can be enlarged arbitrarily by setting $\ell$ sufficiently large.
This property comes from the fact that $ \|h_v(i)\|/\|v\|\rightarrow0$ as $\ell\rightarrow \infty$. 

Note that Verd\'{u}-Wei scheme and Kurosawa-Yoshida scheme cannot use common randomness because  $v$ must be selected in ${\cal V}_i$, which depends on $i$, in their schemes.
Although Moulin-Koetter scheme can use common randomness, the improvement of coding rate is upper bounded by 2 because the codeword $(v, c_v(i))$ of their scheme must satisfy  $\|v\|=\|c_v(i)\|$.
Hence, Coding Scheme 1 is much more efficient than the known coding schemes when common randomness can be used.

\subsection{$K$-MOID Coding with Passive Feedback} \label{sec-2-E}

It is shown in \cite{ad1989-2} that if we can use a passive noiseless feedback channel such that the encoder can know the channel output $Y_t$ at each time $t=1, 2, \cdots, n-1$, the following coding rate can be achieved.
\begin{align}
&\max_{x\in{\cal X}} H(W(\cdot|x))\quad  \mbox{if the encoder is deterministic.} \label{eq-27}\\
&\max_{P\in {\cal P}({\cal X})}H(P\cdot W) \quad\mbox{if the encoder is stochastic.}  \label{eq-28}
\end{align}
Here $W(\cdot|\cdot)$ is the transition probability of the forward channel $W$,  ${\cal P}({\cal X})$ is the set of input probability distributions, and $P\cdot W$ is the output probability distribution for input probability distribution $P\in {\cal P}({\cal X})$.

The above coding rates, \eqref{eq-27} and \eqref{eq-28}, can be achieved by  Coding scheme 1 for $K$-MOID coding as follows. 
We first send $x^{\tilde{n}}$, where $x_t$, $t=1, 2, \cdots, \tilde{n}$, is the optimal fixed input $\tilde{x}$ that achieves the maximum of \eqref{eq-27} in the deterministic case, or is generated by the optimal input probability distribution $\tilde{P}$ that 
achieves the maximum of \eqref{eq-28} in the stochastic case. 
Then the encoder and decoders can obtain random number $v$ from the corresponding channel output $y^{\tilde{n}}$ by using the interval algorithm for random number generation \cite{HH97}.
After $v$ is obtained at the encoder and decoders,  the encoder sends $(h_v(i_1), h_v(i_2), \cdots$, $h_v(i_M ))$ by a transmission code with code length $n^*=Km/r$.

In order to obtain $v$ uniformly distributed over $\{0, 1, 2, \cdots, 2^{\ell m}-1\}$ by the interval algorithm, we use variable $\tilde{n}$. 
Then the expected length ${\rm E}[\tilde{n}]$ is bounded as follows \cite[Theorem 3]{HH97}.
\begin{align}
\frac{\ell m}{H} \leq {\rm E}[\tilde{n}] \leq \frac{1}{H}\left(
\ell m + \log 2(|{\cal Y}|-1)+\frac{h(p_{\max})}{1-p_{\max}}\right),
\end{align}
where $\displaystyle{p_{\max}=\max_{y\in{\cal Y}} P_Y(y)}$, $h(\cdot)$ is the binary entropy function, 
and $H=H(W(\cdot|\tilde{x}))$ or $H=H(\tilde{P}\cdot W)$ if the encode is deterministic or stochastic, respectively.

In this case, coding rate $R$, which is defined by $R=(\log\log N)/({\rm E}[ \tilde{n}]+n^*)$, satisfies that
\begin{align}
R &= \frac{\log \log N}{{\rm E}[ \tilde{n}]+n^*} \nonumber\\
& = \frac{(\ell-3) m + \log (\ell-2) + \log m}{{\rm E}[ \tilde{n}]+Km/r} \nonumber\\
& \rightarrow H \quad \mbox{as $m \rightarrow \infty$ and $\ell\rightarrow \infty$}
\end{align}
where the second equality holds from \eqref{eq-15}, $t=k-1=\ell-3$, and $n^*=Km/r$.

\subsection{MOID Coding with variable $K$} \label{sec-2-F}

In the above, we assumed for simplicity that $K$ is fixed and known. 
But, if $K$ is variable and the decoders don't know $K$, the encoder must send the information of $K$ to the decoders. For instance, this can be realized if we define the encoder $\varphi$ as $\varphi({\cal K}, v)= f(K, v, h_v(i_1), h_v(i_2), \cdots, h_v(i_K))$ instead of \eqref{eq-12}.

If the maximum value of $K$, $K_{\max}$, is given, $K$ can be represented by $\lceil \log K_{\max}\rceil$ bits.
If $K_{\max}$ is not known, $K$ can be represented by Elias $\delta$ code \cite{E1975}, the length of which is not larger than $1+\log K + 2\log(1+\log K)$ bits. 
Since these additional bits can be ignored compared with $n_0=(\ell+K)m$ as $m\rightarrow\infty$, Theorem \ref{thm-1} still holds even if $K$ is variable.
However, we note from \eqref{eq-20} that $\log \log N\approx (\ell -3)m$. Hence, 
$K$ must satisfy that $\log K \ll =n_0=(\ell+K)m = \log\log N -(K-3)m < \log \log N$, which means 
\begin{align}
\lim_{m\rightarrow\infty} \frac{K}{\log N}=0.
\end{align}
Furthermore,  from \eqref{eq-14}, $R$ and $E_2$ decrease to zero as $K$ becomes large for fixed $r$ and $\ell$.

\section{MOID code with ranking} \label{sec-3}
\subsection{Definition of RMOID codes}
In Section \ref{sec-2}, we assumed that selected $K$ receivers are not ranked. But, in this section, we consider the case that $K$ receivers are ranked.
Let $\bK\equiv(i_1, i_2, \cdots,$ $i_K)$, where $i_j$ stands for the receiver of rank $j$.
Then, encoder $\tilde{\varphi}$ and
decoder $\tilde{\psi}_i$ for $K$ ranked receivers can be defined as follows.
\begin{align}
\tilde{\varphi} &: \tilde{{\cal Z}} \times {\cal V} \rightarrow {\cal X}^n \label{eq-3-1} \\
\tilde{\psi}_i &: {\cal Y}^n \rightarrow  \{1, 2, \cdots, K, {\rm F}\}, \label{eq-3-2}
\end{align}
where $\tilde{{\cal Z}}=\{\bK\}$, which is the set of all possible $\bK$, and ${\rm F}$ means ``outside of the ranking".  We call this code $K$-RMOID (ranked-multiple-object identification) code.

Although we can consider many types of errors for this $K$-RMOID code $(\tilde{\varphi}, \tilde{\psi}_1, \tilde{\psi}_2, \cdots, \tilde{\psi}_N)$,
we group the errors into only two types.
To simplify notation, we treat F as rank $K+1$.
Then, the type I (resp. II) error is defined as the error such that a decoded rank of a receiver is larger (resp. smaller) than the true rank of the receiver. 

Let $\tilde{\lambda}_1^{(n)}$ and $\tilde{\lambda}_2^{(n)}$ be the worst probability of
type I and II errors, respectively. Then, they can be represented as follows.
\begin{align}
 \hspace*{-0.2cm} \tilde{\lambda}_1^{(n)}(i_j|\bK)
&\equiv{\rm Pr}\{\tilde{\psi}_{i_j}(\tilde{\varphi}(\bK, V)) > j\} \label{eq-3-3-0}\\
  \tilde{\lambda}_1^{(n)}&\equiv \max_{\bsK\in \tilde{{\cal Z}}}\max_{i_j} 
   \tilde{\lambda}_1^{(n)}(i_j|\bK),   \label{eq-3-3-1}\\
 \tilde{\lambda}_2^{(n)}(i_j|\bK)&\equiv{\rm Pr}\{\tilde{\psi}_{i_j}(\tilde{\varphi}(\bK, V))< j\},\label{eq-3-5}\\
 \tilde{\lambda}_2^{(n)}&\equiv \max_{\bsK\in \tilde{{\cal Z}}}\max_{i_j} 
   \tilde{\lambda}_2^{(n)}(i_j|\bK).   \label{eq-3-5-1}
\end{align}

Furthermore, the error exponents of $\tilde{\lambda}_1^{(n)}$ and $\tilde{\lambda}_2^{(n)}$ are defined by 
\begin{align}
  \tilde{E}_1^{(n)}& \equiv-\frac{1}{n}\log \tilde{\lambda}_1^{(n)},\label{eq-3-4}\\
  \tilde{E}_2^{(n)}& \equiv-\frac{1}{n}\log \tilde{\lambda}_2^{(n)}. \label{eq-3-5-2}
\end{align}

\vspace{0.2cm}
\begin{remark}
From the definition of decoder $\tilde{\psi}_i$ given by \eqref{eq-3-2}, we note 
that $\tilde{\lambda}_1^{(n)}(i_{K+1}|\bK)=\tilde{\lambda}_2^{(n)}(i_1|\bK)=0$.
This means that we can exclude receivers with rank $j=K+1$ (i.e.~F) and the receiver with rank $j=1$ in the maximization $\displaystyle{\max_{i_j}}$ of \eqref{eq-3-3-1} and \eqref{eq-3-5-1},
respectively.
Hence, we can easily check that the type I and II errors defined in this section coincide with the ordinary ones in the case of $K=1$. 
Furthermore, if all ranks $j$, $1\leq j \leq K$, are treated as the same rank, \eqref{eq-3-5}
and \eqref{eq-3-5-1} coincide with \eqref{eq-5-1} and \eqref{eq-6-1}, respectively. Therefore, the 
definition of type I and II errors given by \eqref{eq-3-3-0}-\eqref{eq-3-5-1} are reasonable.
\end{remark}
\vspace{0.2cm}

A triplet $(R, \tilde{E}_1, \tilde{E}_2)$ is said to be achievable by a coding scheme if the following inequalities can be satisfied by the coding scheme.
\begin{align}
\liminf_{n\rightarrow \infty} R_M^{(n)}&\geq R \label{eq-3-6}\\
\liminf_{n\rightarrow \infty} \tilde{E}_1^{(n)} &\geq \tilde{E}_1 \label{eq-3-7}\\
\liminf_{n\rightarrow \infty} \tilde{E}_2^{(n)} &\geq \tilde{E}_2 \label{eq-3-8}
\end{align}

\subsection{Construction of RMOID codes}

For $\bK=(i_1, i_2, \cdots,$ $i_K)$, we define a code $(\tilde{\varphi}, \tilde{\psi}_1, \tilde{\psi}_2, \cdots$, $\tilde{\psi}_N)$ as follows.
\vspace{0.2cm}
\begin{code}\label{code-2}
\begin{align}
\hspace*{-0.5cm}\tilde{\varphi}(\bK, v) \equiv &f(v, h_v(i_1), h_v(i_2), \cdots, h_v(i_K)) \label{eq-3-9}\\
\hspace*{-0.5cm}\tilde{\psi}_i(y^n)
\equiv&\left\{
\begin{array}{ll}
j,  & \mbox{if $h_{\hat{v}}(i)\ne\beta_{l}$, $l=1, 2, \cdots, j-1$}\\
    & \hspace{0.4cm}\mbox{and $h_{\hat{v}}(i)=\beta_j$} \\
{\rm F}, & \mbox{if $h_{\hat{v}}(i)\ne\beta_{l}$, $l=1, 2, \cdots, K$} 
\end{array}\right.\nonumber \\
&\quad\mbox{for $(\hat{v}, \beta_1, \beta_2, \cdots, \beta_M)= g(y^n)$} \label{eq-3-10}
\end{align}
\end{code}
The encoder $\tilde{\varphi}$ is the same as the encoder $\varphi$ of Coding Scheme 1
defined in \eqref{eq-12}. But the order of $h_v(i_j)$ in $f$ of $\tilde{\varphi}$ represents the rank of receiver while the order of $h_v(i_j)$ has no meaning in the case of $\varphi$
defined in \eqref{eq-12}.

As shown in \eqref{eq-3-10}, each decoder $\tilde{\psi}_i$ first checks whether or not receiver $i$ is rank 1. If so, $\tilde{\psi}_i$ outputs 1. Otherwise $\tilde{\psi}_i$ next checks whether or not receiver $i$ is rank 2. If so, $\tilde{\psi}_i$ outputs 2. Otherwise $\tilde{\psi}_i$ checks whether or not receiver $i$ is rank 3. This procedure repeats until rank becomes $K$. Finally, if receiver $i$ is not rank $K$, $\tilde{\psi}_i$ outputs F .

This code $(\tilde{\varphi}, \tilde{\psi}_1, \tilde{\psi}_2, \cdots, \tilde{\psi}_N)$ 
satisfies the following theorem.

\vspace{0.2cm}
\begin{theorem}\label{thm-2}
The following triplet is achievable by Coding Scheme 2 for $K$-RMOID coding.
\begin{align}
(R, &E_1, E_2)=\nonumber\\
&\left(\left(1-\frac{M+3}{M+\ell}\right) r, E(r), \min\left\{\frac{r}{M+\ell}, E(r)\right\}\right),  \nonumber\\
&\hspace{3cm}0 \leq r \leq C, \quad\ell =3, 4, 5, \cdots \label{eq-3-11}
\end{align}
\end{theorem}

\vspace{0.2cm}
{\it Proof} \quad
First we consider the case of the noiseless channel.
For each rank $j$, $j=1, 2, 3, \cdots, K$, 
$\tilde{\lambda}_1^{(n)}(i_j|\bK)$ can be evaluated as follows.
\begin{align}
\tilde{\lambda}_1^{(n)}(i_j|\bK) &=\text{Pr}\left\{\bigcap_{l=1}^{j} \left(h_V(i_j)\ne h_V(i_l)\right)\right\}\nonumber\\
&=0, \label{eq-3-12}
\end{align}
where the last equality holds because $h_V(i_j)=h_V(i_l)$ is satisfied at $l=j$.

Next we derive an upper bound of $\tilde{\lambda}_2^{(n)}(i_j|\bK)$
for receiver $i_j$ with rank $j$.
\begin{align}
\tilde{\lambda}_2^{(n)}(i_j|\bK) &=\text{Pr}\left\{\bigcup_{l=1}^{j-1} \left(h_V(i_j)=h_V(i_l)\right)\right\}\nonumber\\
&\leq \sum_{l=1}^{j-1}
\text{Pr}\left\{h_V(i_j)=h_V(i_l)\right\}\nonumber\\
&\leq {\varepsilon}(j-1)\leq {\varepsilon}K,  \label{eq-3-13}
\end{align}
where the second inequality can be proved in the same way as \eqref{eq-22}.

$\tilde{\lambda}_1^{(n)}(i_j|\bK)$ and the bound of $\tilde{\lambda}_2^{(n)}(i_j|\bK)$
are the same as $\lambda_1^{(n)}(i|{\cal K})$ and the bound of $\lambda_2^{(n)}(i|{\cal K})$ treated in Section II, respectively.
This means that the lower bounds of $\tilde{E}_1^{(n)}$ and $\tilde{E}_2^{(n)}$ 
are the same as the lower bounds of $E_1^{(n)}$ and $E_2^{(n)}$ derived in Section II,
respectively.
Hence, if $(R, E_1, E_2)$ is achievable for code $(\varphi, \psi_1, \psi_2, \cdots, \psi_N)$,
it is also achievable for code $(\tilde{\varphi}, \tilde{\psi}_1, \tilde{\psi}_2, \cdots, \tilde{\psi}_N)$. Therefore, Theorem 2 holds from Theorem 1.
\begin{flushright}Q.E.D.\end{flushright}

\vspace{0.2cm}
\begin{corollary}\label{cor-2}
The $K$-RMOID code constructed by Coding Scheme 2 can attain 
\begin{align}
\lim_{n\rightarrow\infty} R^{(n)} =C, \label {eq-c2-1}\\
\lim_{n\rightarrow\infty} \tilde{\lambda}_1^{(n)} =0, \label{eq-c2-2}\\
\lim_{n\rightarrow\infty} \tilde{\lambda}_2^{(n)} =0.\label{eq-c2-3}
\end{align}
\end{corollary}

\vspace{0.2cm}
{\it Proof} \quad Corollary \ref{cor-2} can be proved in the same way as Corollary \ref{cor-1}.
\vspace{-0.4cm}\begin{flushright}Q.E.D.\end{flushright}

\vspace{0.2cm}
\begin{remark}
The same arguments treated in Sections \ref{sec-2-C} to \ref{sec-2-F} also hold 
 for $K$-RMOID code $(\tilde{\varphi}, \tilde{\psi}_1, \tilde{\psi}_2, \cdots$, $\tilde{\psi}_N)$.
\end{remark}

\section{Conclusion}
In this paper, we defined the MOID coding and we proposed efficient explicit MOID coding schemes for non-ranked and ranked cases. 
We also considered the MOID coding with common randomness, noiseless passive feedback,  transmission coding, and variable $K$ coding.

Although we don't consider the converse part of the coding theorem for the MOID coding, 
it is an interesting open problem.

\ifCLASSOPTIONcompsoc
\else
\fi

\ifCLASSOPTIONcaptionsoff
  \newpage
\fi


\begin{thebibliography}{9}
\bibitem{ad1989} 
R.~Ahlswede and G.~Dueck, 
``Identification via channels," 
{\em IEEE Transactions on Information Theory}, vol.~35, no.~1, pp.~15--29, Jan.~1989.
\bibitem{hv1992}
T.~S.~Han and S.~Verd\'{u}, ``New result in the theory of identification via channels," 
{\em IEEE Transactions on Information Theory}, vol.~38, no.~1, pp.~14--25, Jan.~1992.
\bibitem{vw1993}
S.~Verd\'{u} and V.~K.~Wei, ``Explicit construction of optimal constant-weight codes for identification via channels,"  {\em IEEE Transactions on Information Theory}, 
vol.~39, no.~1, pp.~30--36, Jan.~1993.
\bibitem{ky1999}
K.~Kurosawa and T.~Yoshida, ``Strongly universal hashing and identification codes via channels," {\em IEEE Transactions on Information Theory}, 
vol.~45, no.~6, pp.2091--2095, June 1999.
\bibitem{mk2006}
P.~Moulin and R.~Koetter, ``A framework for the design of good watermark identification codes," {\em SPIE Proceedings 6072}, Security, Steganography, and Watermarking of Multimedia Contents VIII, 
pp.~60721H-1--60721H-10, Jan.~2006. 
\bibitem{ahlswede2006} 
R.~Ahlswede, ``Introduction," General theory of information transfer and combinatorics, LCNS4123, Springer, pp.~1-44, 2006
\bibitem{ahlswede2008}
R.~Ahlswede, ``General theory of information transfer: Updated,"
Discrete Applied Mathematics, Elsevier, vol.~156, pp.~1348--1388, 2008.
\bibitem{abk2006} R.~Ahlswede, B.~Balkenhol, and C.Kleinew\"{a}chter,
``Identification for sources," General theory of information transfer and combinatorics, LCNS4123, Springer, pp.~51-61, 2006
\bibitem{ad1989-2}
R.~Ahlswede and G.~Dueck, 
``Identification in the Presence of Feedback -- A Discovery of New Capacity Formulation," 
{\em IEEE Transactions on Information Theory}, vol.~35, no.~1, pp.~30--36, Jan.~1989.
\bibitem{HH97} T.S.~Han and M.~Hoshi, ``Interval Algorithm for Random Number Gerenation,"
{\em IEEE Transactions on Information Theory}, vol.~43, no.~2, pp.~599--611, March 1997.
\bibitem{E1975}
P.~Elias, ``Universal codewords sets and representations of the integers,"
{\em IEEE Transactions on Information Theory}, 
vol.~IT21, no.~2, pp.~194--203, March 1975
\bibitem{YU2014}
H.~Yamamoto and M.~Ueda, ``Identification codes to identify multiple objects,"
2014 IEEE International Symposium on Information Theory, pp.~1241--1245, 2014

\end{thebibliography}
\end{document}